# The Ground State of the Pseudogap in Cuprate Superconductors

T. Valla,[1*] A. V. Fedorov,[2] Jinho Lee,[3] J.C. Davis[3] and G. D. Gu[1]

[1]Condensed Matter Physics and Materials Science Department, Brookhaven National Laboratory, Upton, NY 11973, USA.

[2]Advanced Light Source, Lawrence Berkeley National Laboratory, Berkeley, CA 94720, USA.

[3]LASSP, Department of Physics, Cornell University, Ithaca, NY 14853 USA

**We present studies of the electronic structure of $La_{2-x}Ba_xCuO_4$, a system where the superconductivity is strongly suppressed as static spin and charge orders or 'stripes' develop near the doping level of $x=1/8$. Using angle-resolved photoemission and scanning tunneling microscopy, we detect an energy gap at the Fermi surface with magnitude consistent with *d*-wave symmetry and with linear density of states, vanishing only at four nodal points, even when superconductivity disappears at $x=1/8$. Thus, the non-superconducting, 'striped' state at $x=1/8$ is consistent with a phase incoherent *d*-wave superconductor whose Cooper pairs form spin/charge ordered structures instead of becoming superconducting.**

* To whom correspondence should be addressed. E-mail: valla@bnl.gov

There are several generally accepted phenomena in high temperature superconductivity (HTSC) that make the cuprates so fascinating. One of them is a *d*-wave symmetry of the superconducting gap. Another feature is a normal state gap ('pseudogap') in underdoped materials, which exists above the temperature of the superconducting transition $T_C$ *(1, 2)*. There are multiple aspects to the pseudogap phenomenon. In particular, the distinction is usually made between the 'large' pseudogap in the overall density of states (DOS), and a 'small' pseudogap in the excitations at the Fermi surface, seen in spectroscopic probes such as angle-resolved photoemission spectroscopy (ARPES) *(1, 2)*. Here, we consider the 'small' pseudogap. It is generally believed and observed that the magnitude of the pseudogap monotonically decreases with increasing doping, whereas $T_C$ moves in the opposite direction in the underdoped regime *(1, 2)*. The origin of the pseudogap and its relationship to superconductivity is one of the most important open issues in physics of HTSC and represents the focal point of current theoretical interest *(3-6)*. In one view, the pseudogap is a pairing (superconducting) gap, reflecting a state of Cooper pairs without global phase coherence. The superconducting transition then occurs at some lower temperature when phase coherence is established *(7)*. In an alternative view, the pseudogap represents another state of matter that competes with superconductivity. However, the order associated with such a competing state has never been unambiguously detected. The first hints came from neutron scattering studies in a magnetic field, where an incommensurate spin order was detected inside vortices *(8)*. However, it was not until recent scanning tunnelling microscopy (STM) experiments that more was learned about any potential candidate for such 'hidden order'. A charge ordered state, energetically very similar to the superconducting state, has been found in the vortex cores *(9)*, in the 'pseudogap' regime *(10)* above $T_C$ and in patches of underdoped material in which the coherent conductance peaks were absent *(11)*. We show that a similar state represents the ground state in a system with strongly suppressed superconductivity and with a static spin *(12)* and charge *(13)* orders - La$_{2-x}$Ba$_x$CuO$_4$ (LBCO) at *x*=1/8. The *k*-dependence of the gap in this state looks the same as the superconducting gap in superconducting cuprates: it has magnitude consistent with *d*-wave symmetry and vanishes at four nodal points on the Fermi surface. Furthermore, the single particle gap, measured at low temperature, has surprising doping dependence with a maximum at x≈1/8, precisely where the charge/spin order is established between two adjacent superconducting domes. These findings reveal the pairing origin of the 'pseudogap' and imply that the most strongly bound Cooper pairs at x≈1/8 are most susceptible to phase disorder and spatial ordering *(7, 14, 15)*.

LBCO exhibits a sharp drop in superconducting transition temperature, $T_C \rightarrow 0$, when doped to ~1/8 holes per Cu site (*x*=1/8) *(16)*, while having almost equally 'strong' superconducting phases, with $T_{Cmax}$≈30 K at both higher and lower dopings. Therefore, the *x*=1/8 case represents an ideal system to study the ground state of the pseudogap as the 'normal' state extends essentially to T=0. In scattering experiments on single crystals, a static local spin order with period of 8 unit cells *(12, 14)* and a charge order *(13)* with period of 4 unit cells – so called 'stripes' - have been detected at low temperatures. While superconductivity is strongly reduced at *x*=1/8, metallic behaviour seems to be preserved. Optical studies have detected a loss of spectral weight at low frequencies with simultaneous narrowing of a Drude component, suggesting the development of an anisotropic gap *(17)*. Here, we use ARPES and STM to measure the electronic excitations and detailed momentum dependence of the single-particle gap in the ordered state of LBCO.

Figure 1 shows the photoemission spectra from LBCO at $x=1/8$ in the ordered state (T=16 K). In panel (A) the momentum distribution of the photoemission intensity from the energy window of ±10meV around the Fermi level is shown within the Brillouin zone. From these and other contours measured for several samples, we extracted the 'Fermi surface' as a line in momentum space that connects the maxima of each of the measured momentum distribution curves (MDC) at $\omega=0$. In addition, we also show the extracted Fermi surface of LSCO at $x=0.07$ that agrees well with published data *(18)*. The areas enclosed by the Fermi lines correspond to $x=0.06\pm0.015$, for LSCO and $x=0.115\pm0.015$, for LBCO, in good agreement with the nominal doping levels, signalling that the bulk property has been probed. In both systems we have detected an excitation gap (*19*) with a magnitude that depends on the $k$ position on the Fermi surface, vanishing at the node and with maximum amplitude near the anti-node as shown in panels (B) and (C).

In the detailed $k$-dependence for several samples (Fig. 2), two unexpected properties are uncovered: first, gaps in all samples have magnitudes consistent with $d$-wave symmetry even though superconductivity is essentially non-existent in LBCO at $x=1/8$. Second, the gap in LBCO is larger at $x=1/8$ than at $x=0.095$ and than in LSCO ($x\approx0.07$). This contradicts a common belief that the excitation gap in cuprates monotonically increases as the anti-ferromagnetic (AF) phase is approached. Fig 2C shows the compilation of the maximal gap values, $\Delta_0$, in LSCO and LBCO systems, as a function of doping, from this study and from previously published work. Values for LSCO for $x=0.063$ and $x=0.09$ are extracted from Fig. 4 in Ref. *18* and those for $x\geq0.1$ are from Ref. *2*. All the points have been measured at T≈20 K: in the superconducting state for $x=0.09, 0.095, 0.1, 0.165$ and $0.22$ and in the 'normal' state for the other samples. It is clear from the figure that the total gap is not monotonic. Rather, in LBCO it peaks at or near $x=1/8$ when superconductivity vanishes and 'stripes' are fully developed.

The momentum resolved picture from ARPES is consistent with the STM data obtained from an LBCO sample at $x=1/8$, cut from the same parent crystal used for ARPES. In Fig. 3A a typical STM topographic image of a cleaved LBCO surface is shown. In addition, the dI/dV spectra were taken at many points in a wide range of energies (Fig S2). In the averaged (over the whole field-of-view in Fig. 3A) conductance spectrum (Fig. 3B), a symmetric 'V'-like shape at low energies, with zero-DOS falling exactly at the Fermi level, consistent with a pairing $d$-wave gap. The magnitude of this gap, $\Delta_0 \approx 20$ meV, as determined from the breaks in dI/dV curve agrees with the maximal gap $\Delta_0$ measured in photoemission.

Our study provides the evidence for a $d$-wave gap in the 'normal' ground state of a cuprate material. Previous studies on underdoped $Bi_2Sr_2CaCu_2O_{8+\delta}$ (BSCCO) were always affected by the superconductivity: the disconnected 'Fermi arcs' were seen, shrinking in length as temperature was lowered below T* and collapsing onto (nodal) points below $T_C$ *(20, 21)*. Due to this abrupt intervention of superconductivity it was not clear if the pseudogap ground state would have a Fermi arc of finite length, a nodal point or if it would be entirely gapped. In LBCO, the absence of superconductivity at $x=1/8$ has enabled us to resolve this puzzle and to show that the 'normal state gap' has isolated nodal points in the ground state. This result points to the 'pairing' origin of the pseudogap, in general agreement with recent thermal transport measurements *(22)*. With increasing temperature, a finite length Fermi arc forms, as suggested in Fig. 2B, in accord with results on BSCCO *(20, 21)*.

What might be the origin of the observed *d*-wave gap in LBCO if superconductivity is absent? Neutron and x-ray scattering studies on the same crystal have identified a static spin order and a charge order *(12, 13)*. Therefore, it would be tempting to assume that at least a portion of the measured gap is due to the charge order, in analogy with conventional 2-dimensional (2D) charge-density-wave (CDW) systems. It has been suggested that in cuprates the spin/charge ordered state forms in a way where carriers doped into the AF insulator segregate into one-dimensional (1D) charge rich structures ('stripes') separated by the charge poor regions of a parent antiferromagnet *(14, 23-25)*. However, questions have often been raised on how to reconcile these unidirectional structures with an apparent 2D Fermi surface and a gap with *d*-wave symmetry. In the more conventional view, doped carriers are de-localized in the planes, forming a 2D Fermi surface that grows in proportion with carrier concentration. The charge/spin ordered state may then be formed in the particle-hole channel by nesting of Fermi surface segments, producing a divergent electronic susceptibility and a Peierls-like instability and pushing the system into a lower energy state with a single particle gap at nested portions of the Fermi surface. An example of a cuprate where such a 'nesting' scenario is proposed to be at play is $Ca_{2-x}Na_xCuO_2Cl_2$ (CNCOC) *(26)*. STM studies have detected checkerboard-like modulations in local DOS on the surface of this material, with $4a \times 4a$ periodicity, independent of doping *(27)*. Subsequent ARPES studies on the same system have shown a Fermi surface with a nodal arc and truncated anti-nodal segments *(26)*. The anti-nodal segments can be efficiently nested by $q_{CDW}=2k_F=\pi/(2a)$ (and $3\pi/(2a)$) – the same wave vectors observed in STM for charge superstructure, making the nesting scenario viable, at least near the surface of CNCOC. However, if we apply the same nesting scenario to LBCO at *x*=1/8, we obtain $q_{CDW} \approx 4k_F$ (=$\pi/2a$), for charge order, instead of $2k_F$ nesting, suggested to be at play in CNCOC. Moreover, the nesting of anti-nodal segments would produce a wave-vector that shortens with doping, opposite of that observed in neutron scattering studies in terms of magnetic incommensurability. This is illustrated in Fig.4 where we compile the doping dependences of several relevant quantities.

There is another, more fundamental problem with the 'nesting' scenario: any order originating from nesting (particle-hole channel) would open a gap only on nested segments of the Fermi surface, preserving the non-nested regions. The fact that only four gapless points (nodes) remain in the ground state essentially rules out nesting as an origin of pseudogap. In addition, a gap caused by conventional spin/charge order would be pinned to the Fermi level only in special cases. The observation that it is always pinned to the Fermi level (independent of $k$-point, as measured in ARPES and of doping level, as seen in STM on different materials) and that it has $d$-wave symmetry undoubtedly points to its pairing origin – interaction in the particle-particle singlet channel (*28*). Note that, in contrast to the low-energy pairing gap, STM at higher energies shows a DOS suppressed in a highly asymmetric manner, indicating that some of the 'nesting'-related phenomena might be at play at these higher energies (Fig. 3B).

The surprising anti-correlation of the low-energy pairing gap and $T_C$ over some region of the phase diagram suggests that in the state with strongly bound Cooper pairs, the phase coherence is strongly suppressed by quantum phase fluctuations. Cooper pairs are then susceptible to spatial ordering and may form various unidirectional *(14, 24, 25)* or 2D *(15, 27-30)* superstructures. Quantum phase fluctuations are particularly prominent in cases where such superstructures are anomalously stable. For some of the proposed structures this occurs at the doping of 1/8, in general agreement with our results: 1/8 represents the most prominent 'magic fraction' for a checkerboard-like 'CDW of Cooper pairs' *(15)*, and it locks the 'stripes' to the lattice in a unidirectional alternative. The presence of nodes in the ground state of the pseudogap represents a new decisive test for validity of models proposed to describe such structures.

**Figure legends:**

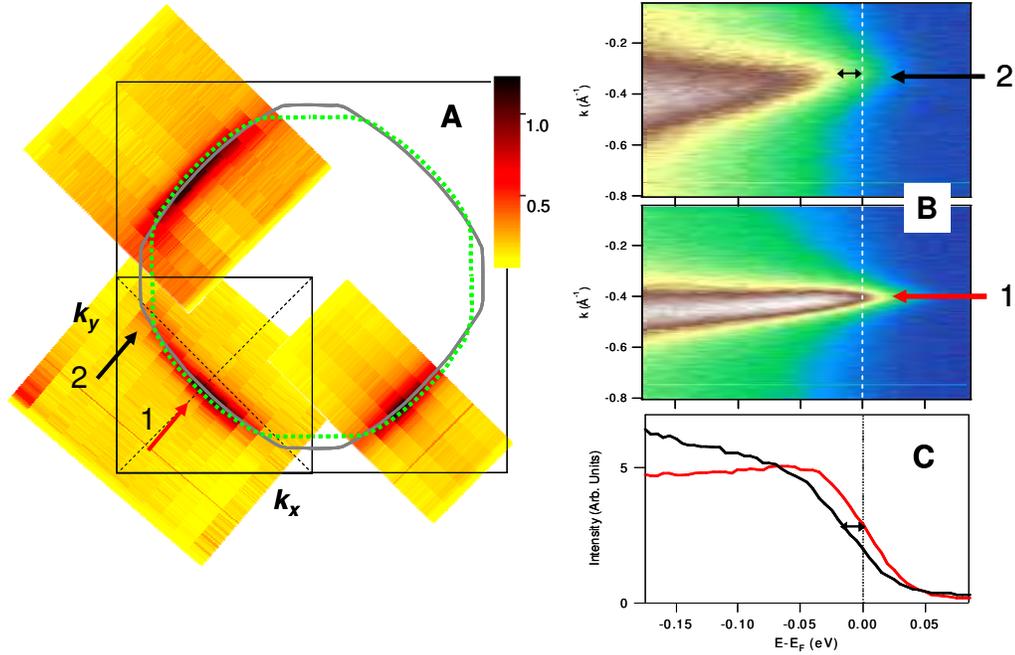

**Fig 1.** Photoemission from LBCO at *x*=1/8. **A)** Photoemission intensity from a narrow interval around the Fermi level ($\omega=0\pm10$ meV) is shown as a function of the in-plane momentum. High intensity represents the underlying Fermi surface. Lines represent fits to the positions of maxima in probed MDCs for LBCO (*x*=0.125) (solid line) and LSCO (*x*=0.07) (dashed line) **B)** Photoemission intensity from LBCO sample as a function of binding energy along the momentum lines indicated in A) by arrows. **C)** Energy distribution curves of spectral intensity integrated over a small interval $k_F\pm\Delta k$ along the two lines in *k*-space shown in B. The arrow represents the shift of the leading edge. The spectra were taken in the charge ordered state at T=16 K.

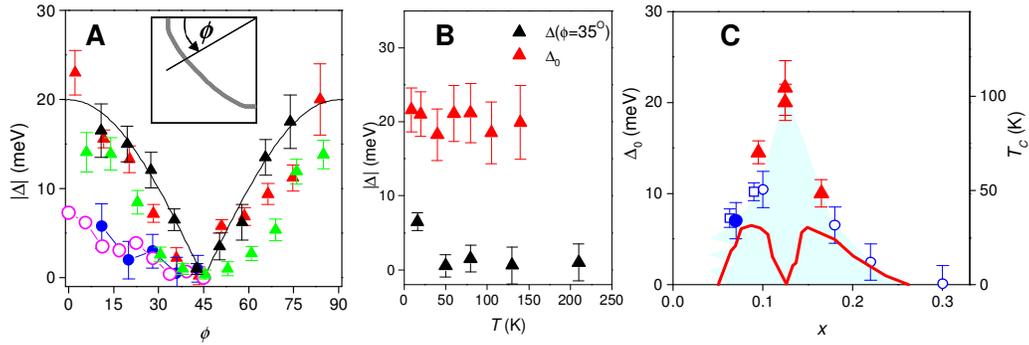

**Fig 2.** *k*- and doping-dependence of the single-particle gap. **A)** Magnitude of single-particle gap (leading edge gap) at T=16 K as a function of an angle around the Fermi surface, as defined in the inset, for LBCO at *x*=1/8 (black and red triangles), *x*=0.095 (green triangles) and LCSO at *x*=0.07 (blue circles) have been measured. Points for LSCO at *x*=0.063 (magenta circles) have been extracted from Fig. 4A from Ref. *18*. The line represents a *d*-wave gap amplitude, $\Delta_0|\cos(2\phi)|$ with $\Delta_0$=20 meV. **B)** Temperature dependence of $\Delta_0$ (red triangles) and $\Delta(\phi\approx35°)$ (black triangles) for LBCO at *x*=1/8. **C)** Doping dependence of $\Delta_0$ in LBCO (triangles) and LSCO (circles and squares). Solid symbols: this study, open squares: Ref. *18*, open circles: Ref. *2*. Red line represents doping dependence of $T_C$ for LBCO from Ref. *16*.

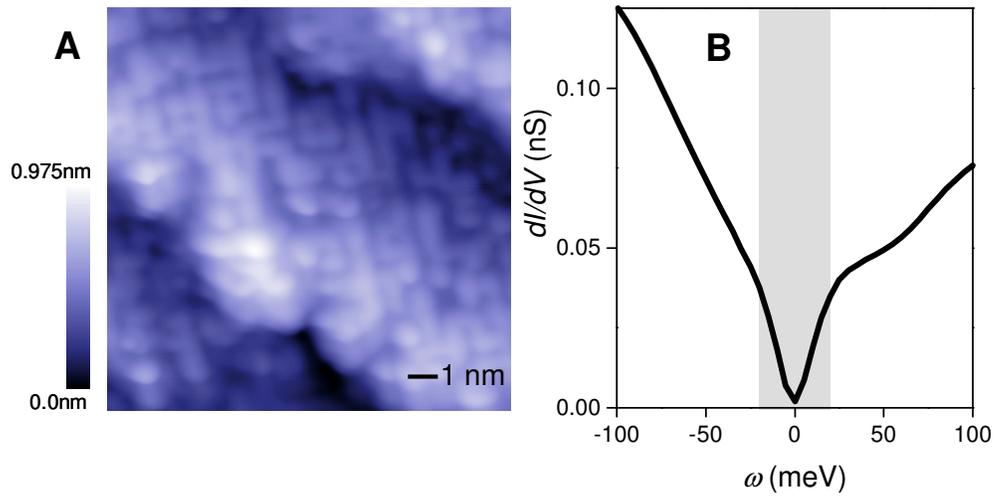

**Fig. 3.** STM of LBCO. **A)** High-resolution STM topographic image of the cleaved sample. The image was taken at 4.2K. **B)** A tunnelling conductance spectrum averaged over the area shown in A). A 'V-like' profile of density of states for energies $|\omega|\leq 20$ meV (gray region) is consistent with a $d$-wave gap observed in ARPES.

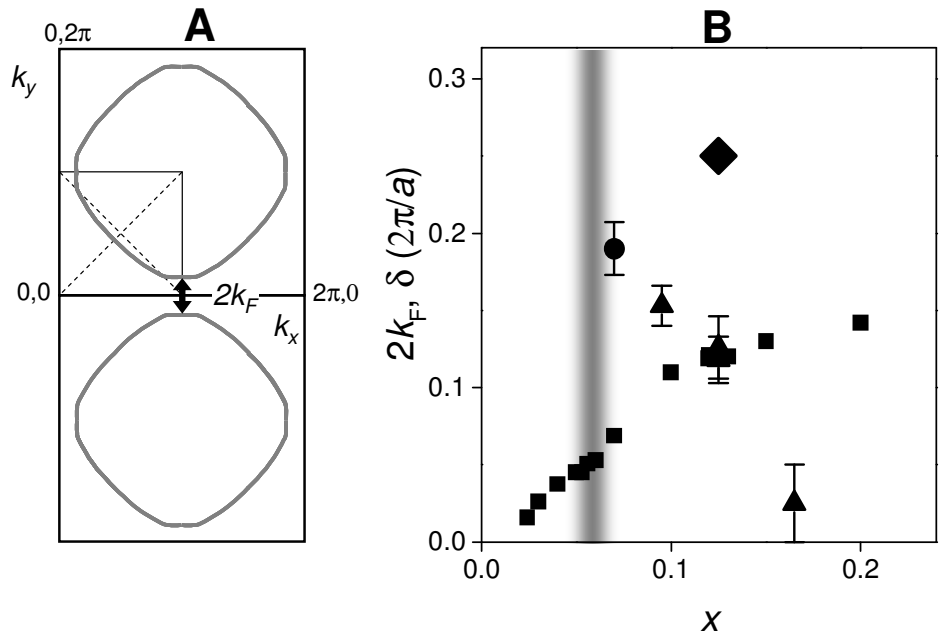

**Fig. 4.** A compilation of relevant wave-vectors from neutron and x-ray scattering and ARPES on 214 materials. **A)** A sketch of relevant vectors in the *k* space. **B)** Doping dependence of the anti-nodal $k_F$, indicated in A) by the arrow, in LBCO (triangles) and LSCO (circles). Also shown are wave-vector for charge order (*13*) (diamond) and the incommensurability $\delta$ from $(\pi,\pi)$ point from neutron scattering experiments (*31*, *32*) (squares). The grey vertical bar represents the boundary between the 'diagonal' and 'parallel' spin superstructures and the onset of superconductivity.

**Supporting Online Material**

www.sciencemag.org

Materials and Methods

Figs. S1, and S2

References

## Supporting Online Material

**Materials and Methods**

**Single crystals** of $La_{2-x}Ba_xCuO_4$ (LBCO) with $x$=0.095, 0.125, and 0.165 and $La_{2-x}Sr_xCuO_4$ (LSCO) with $x$=0.07 were grown by the traveling solvent floating zone method at Brookhaven National Laboratory. The superconducting transition temperatures, determined by magnetic susceptibility, were 32K for $x$=0.095 and 24 K for $x$=0.165 LBCO samples, while the LSCO sample had $T_C \approx 20$ K. The LBCO sample at $x$=1/8 had a strongly suppressed bulk $T_C \approx 2.4$ K and a small volume fraction (<1%) of a component with higher transition temperature, $T_C \sim 20$K.

**Angle-Resolved Photoemission Spectroscopy** (ARPES) experiments were performed at the beamline 12.0.1 of the Advanced Light Source with Scienta SES-100 electron spectrometer operating in the angle-resolved mode. Samples were mounted on a high precision variable-temperature goniometer with three angular and three linear degrees of freedom and cleaved at low temperature (~15-20 K) under ultra-high vacuum (UHV) conditions (~$1 \times 10^{-9}$ Pa). The photon energy was tuned to ~110 eV, and the overall energy resolution was set to ~35 meV.

The gap is extracted as a "leading edge" position or the inflection point of the energy distribution curve (EDC) of spectral intensity integrated over a small interval $k_F \pm \Delta k$ along the measured lines in $k$-space. The integration interval ($\Delta k$=0.08 Å$^{-1}$) was wide enough to bring the "leading edge" of the nodal line EDC to the Fermi level. Without the integration, the leading edge of any gapless spectral function with a peak at zero energy is always shifted slightly above the Fermi level. The shift depends primarily on the energy resolution, but $k$-resolution, temperature, width of the peak and the velocity of a state also play a role. In our case, the leading edge of the nodal spectral function (no integration) was ~ 3 (5) meV above the $E_F$ for the spectra taken with 25 (35) meV resolution, whereas in the anti-nodal region, the leading edge position was completely insensitive to the width of the integration interval for $0<\Delta k<0.1$ Å$^{-1}$. The gap magnitudes obtained by integration around $k_F$ are more accurate than those extracted without the integration as they do not contain the false contribution from the gapless nodal region. When compared to data from the literature (*S1*, *S2*), one should keep in mind that those gap magnitudes were obtained without integration. Our simulations of the nodal spectra show that the somewhat better energy resolution (15-20meV) in those studies makes a small contribution (~2 meV) to the total nodal-antinodal difference in the leading edge position, similar to what was observed in Ref. (*S2*), Fig. 66.

Photoemission spectra were recorded in a geometry where the polarization of light was perpendicular to the probed momentum line. The spectra used to map the k-space and to extract the gap were recorded along the lines nearly parallel to the nodal line. Such a geometry is favourable for emission from the nodal region, but less favourable for emission from the anti-nodal region. This disparity could, in principle, affect the measured gap magnitudes. In Fig. S1 we show that the leading edge position is not sensitive to the polarization of light. The same point on the Fermi surface near the anti-node was recorded both in the geometry that favours nodal emission and in the geometry that favours emission from the anti-nodal region. Although there is a difference in the line-shape, the measured gap remains unaffected.

**Scanning Tunnelling Microscopy** (STM) measurements were performed using low temperature STM located in the ultra-low vibration laboratory of Cornell University. Samples of LBCO were cut in ~ 1mm size after determining the orientation of the *c-* axis, and attached to the sample holder with conductive epoxy. Samples were cryogenically cleaved *in-situ* in similar conditions to the ARPES measurements, and inserted into the STM head. Etched tungsten tips, cleaned by field emitting on a Au single crystal in UHV were used. The typical junction resistance used in the STM measurements on LBCO was around 100 GΩ for topography, and 5 GΩ for spectroscopy. The differential conductance or *dI/dV* spectra were measured up to ± 0.5V on a 128 × 128 grid over ~ 13nm × 13nm field of view (FOV), using a lock-in technique with 5mV resolution. As shown in Fig. S2, the conductance varies significantly over the measured FOV, indicating surface inhomogeneity. Some regions show only broad humps at high energies with diminishing density of states at low energies, while the others (~3% of the FOV area) show spectra with coherence peaks at energies $|\omega|$~10-15 meV, probably representing a small volume fraction of superconducting component with higher $T_C$ that was also detected in magnetic susceptibility measurements. Fig. 3B is the average *dI/dV* over ~16,000 spectra, i.e. the whole FOV.

Both ARPES and STM can only measure the magnitude of the order parameter $|\Delta(k)|$.

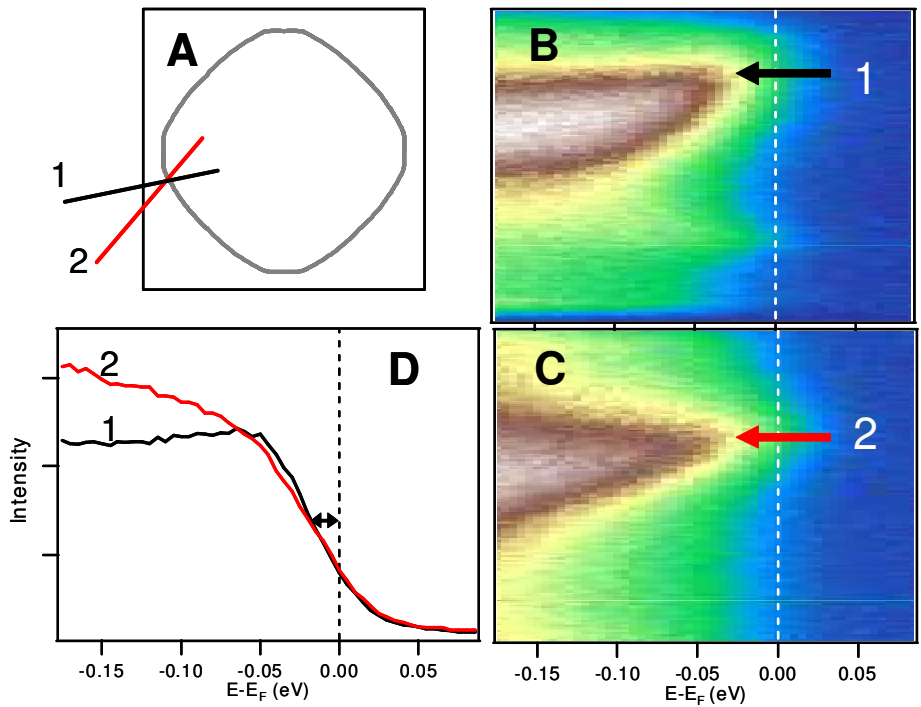

**Fig. S1. A)** A schematic view of Brillouin zone and the Fermi surface of LBCO at x=1/8. Lines (1) and (2) correspond to the momentum lines probed in ARPES in panels B and C. **B)** and **C)** contours of photoemission intensity along the two momentum lines indicated in A. **D)** Energy distribution curves of spectral intensity integrated over a small interval $\Delta k$ around the same $k_F$ along the two measured lines in B and C. The arrow represents the shift of the leading edge.

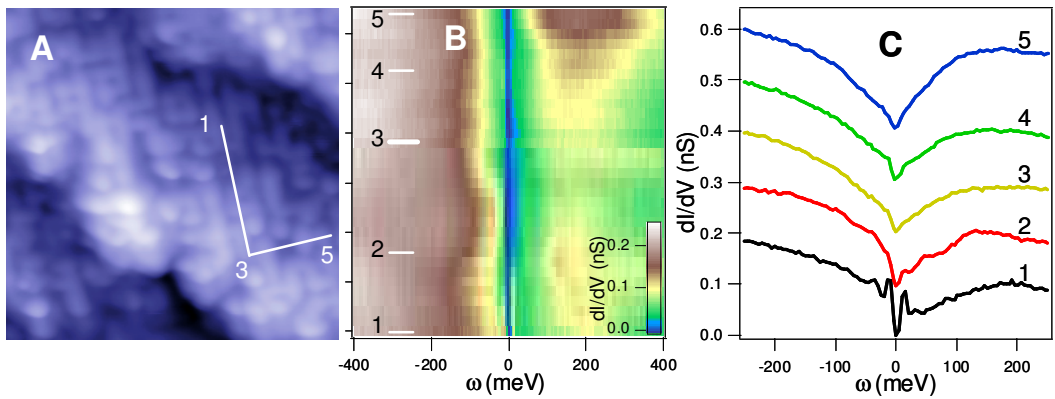

**Fig. S2. A)** High-resolution STM topographic image of the cleaved sample taken at 4.2K. **B)** A contour of tunnelling conductance recorded along the two lines indicated in A. **C)** Conductance spectra at several points along the path 1-3-5 as indicated in A and B.